\documentclass[preprint,floats,superscriptaddress,pre]{revtex4}

\expandafter\let\csname equation*\endcsname\relax

\expandafter\let\csname endequation*\endcsname\relax

\usepackage{amsmath}

\usepackage{graphicx,psfrag,bbm,latexsym,color,dcolumn,bm,dsfont,bbm,
color,mathrsfs,bbold,latexsym,amsfonts,amssymb}

\usepackage{hyperref} 

\usepackage{amssymb}
\newcommand{\media}[1]{\left\langle #1 \right\rangle}

\definecolor{myred}{RGB}{168,5,14}

\definecolor{editorcolor}{RGB}{0,0,0}

\begin{document}

\title
{Exact results for the Ising model on a small-world network}

\author{M. Ostilli}
\affiliation{Instituto de F\'isica,  Universidade Federal da Bahia, Salvador--BA, 40170-115, Brazil}


\begin{abstract}
  Small-world networks provide an interesting framework for studying the interplay between regular and random graphs,
  where links are located in a regular and random way, respectively. 
  On one hand, the random links make the model to obey some kind of mean-field behavior. On the other hand, the links of
  the regular lattice make the system to retain some related non trivial correlations. The coexistence of these two
  features in general prevent a closed analytical treatment. Here we consider a one-dimensional small-world Ising model and derive analytically its
  equation of state, critical point, critical behavior, and critical correlations.
  Despite being one of the simplest small-world models, our exact and intuitive analysis reveals some intriguing properties.
\end{abstract}

\maketitle

\email{massimo.ostilli@gmail.com}

\section{Introduction}
A quarter of a century ago, Watts and Strogatz~\cite{Watts} introduced a very interesting network model, later on called the small-world (SW) network,
which was able to interpolate between regular lattices and random networks. More precisely, by changing a single parameter $p\in [0,1]$ 
related to the probability of random connections between nodes, SW networks can be used for 
studying the interplay between regular short-range (SR), i.e., finite-dimensional systems, and random long-range (LR), i.e., infinite-dimensional systems.
Since then, SW networks have attracted a lot of attention and many different graph versions, as well as many approaches to study statistical models running on the
top of such graphs, have been proposed. The most common and peculiar feature of all these models is that, in the thermodynamic limit (TDL), for any value of $p$ in $(0,1]$,
due to the SW effect, by which the mean distance between nodes scales only logarithmically with the number of nodes $N$,
  they obey some kind
of mean-field behavior. In particular, the Ising model built on the top of SW graphs having simple degree distributions (i.e., not fat-tailed), shows the classical
Curie-Weiss mean-field behavior. On the other hand, the same model, for any value of $p$ in $[0,1)$, also retains some non trivial correlations of the
pure finite-dimensional lattice, i.e., the model before applying the random connection procedure (where $p=0$).
The coexistence of these two opposite features, in general, prevents an analytical approach. Indeed, for the Ising model, exact results seem partial and often
confined to specific SW models~\cite{Skantzos}-\cite{SW2}.
Before making a very short review of these results, all related to the Ising model, we recall that there exist mainly two ways to attain a SW graph by starting from
a regular lattice: one consists in rewiring at random the SR links passing through each node with probability $p$ (shortly, ``rewired SW''), the other consists in adding
$cN/2$ random links (shortly, ``added SW''), $c$ being the additional connectivity. Within each of these two categories, however, there are still several variants.
In ~\cite{Skantzos}, in the framework of neural networks, the phase diagram of the annealed version of the $d$-dimensional added SW model is obtained.
In ~\cite{Barrat}, by using the replica method, a simplified one-dimensional rewired SW model is exactly solved. 
In~\cite{Hastings}, by using Gaussian transform and scaling arguments, the critical behavior of the annealed version of the $d$-dimensional added SW model is obtained.
In~\cite{Viana}, by using combinatorial arguments, a specific one-dimensional added SW model,
where random links connect only equally spaced nodes of the one-dimensional regular lattice, is exactly solved.
In ~\cite{Niko} and~\cite{Niko2}, on the base of the replica method, an analytical approach has
been developed for the one-dimensional added SW model,
but, due to the degree fluctuations, the calculations here are rather involved and require a numerical support for evaluating the effective field distributions.
In ~\cite{Berker}, exact results on a hierarchical added SW are obtained.
In ~\cite{Hastings2} and in ~\cite{SW}, by using series expansions, the exact critical temperature of the $d$-dimensional added SW model is found;
including the case for disordered and frustrated couplings for the spin-glass phase transition~\cite{SW,SW2}.

Here, we consider an Ising model built (quenched) on a one-dimensional added SW model where the LR links and the SR links are associated to two independent couplings. More precisely,
the SR links define a regular one-dimensional ring, while the LR links are disposed at random and in such a way that each node
is attached to exactly $k$ LR links. In other words, the resulting one-dimensional SW graph is obtained by superimposing
a regular random graph (RRG)~\cite{Bollobas,Note} of degree $k$ onto a one-dimensional ring.
The few loops of the RRG (having length which scale as $\log(N)$), as well as the single loop of the one-dimensional ring, become negligible in TDL, allowing therefore
for an exact application of the Belief-Propagation (BP) method (also known under the names of Cavity-Method, Bethe-Peierls approach, or Message-Passing-Algorithm)
~\cite{Yedidia,Yedidia2,Mezard_Parisi,Mezard_Montanari}. 
In the BP approach, messages flowing through the links reach a stationary distribution which in turn determines the thermodynamic quantities of interests, like
mean magnetization and correlations.
Moreover, as we shall see, the absence of fluctuations of the node degrees implies the absence of fluctuations of the stationary messages, i.e., their
stationary distribution is a delta-Dirac pecked around two homogeneous values, one for the SR link, and one for the LR link. 
As we shall see, this fundamental simplification allows to determine in a closed analytical manner the equation of state, the critical temperature and to rigorously
derive the critical behavior together with its non universal quantities which, quite interestingly, for certain values of the SR and LR couplings,
own a singularity. We also evaluate the correlations near the critical point and draw some conclusions.
To the best of our knowledge, within one-dimensional SW Ising models, the proposed exact solution provides the most comprehensive and compact-intuitive approach to the problem. 

\section{Model definition and the BP approach}
\begin{figure}
  \begin{center}
    {\includegraphics[height=8cm,clip]
      {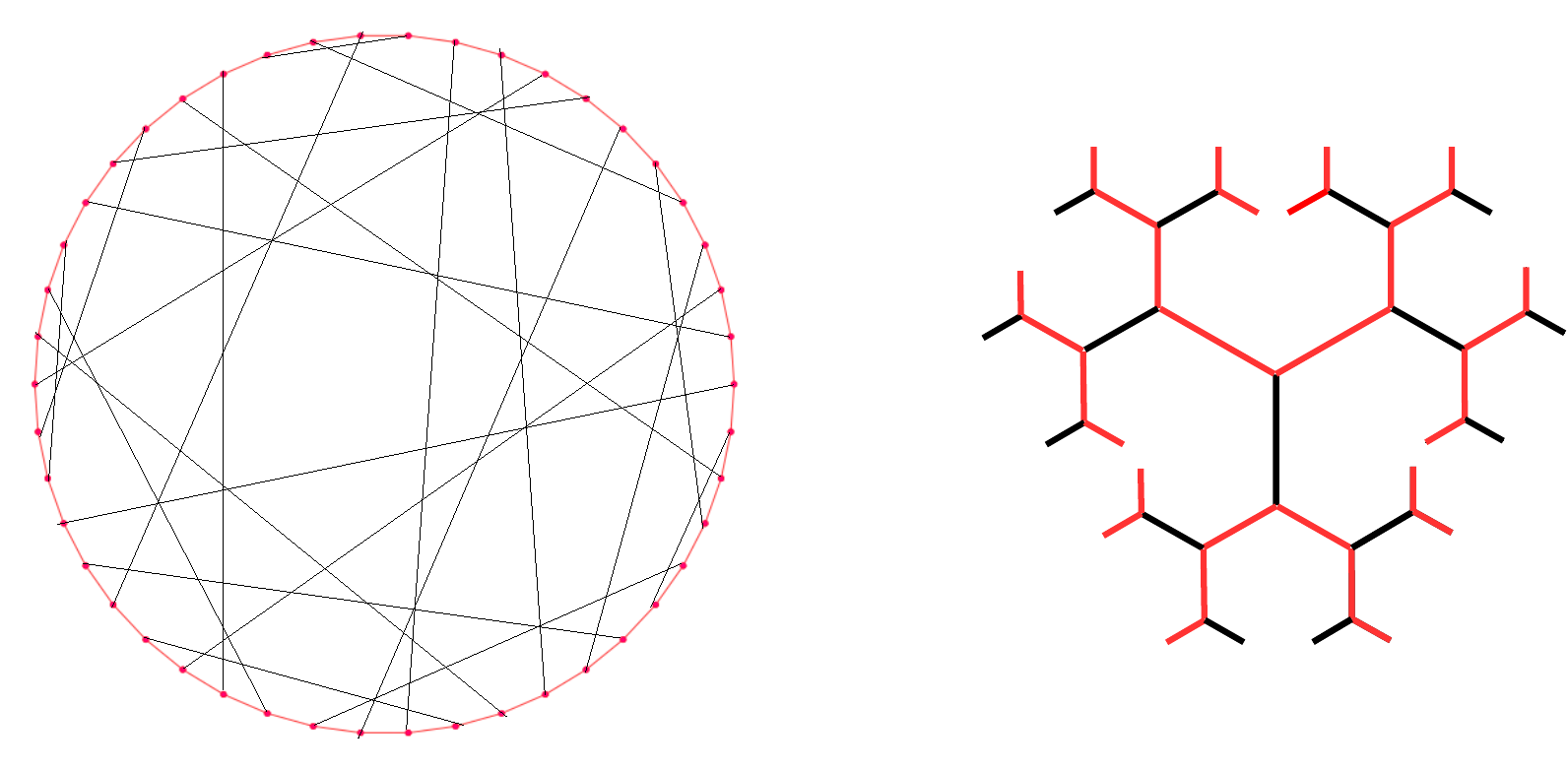}}
    \caption{Left: the one-dimensional SW graph discussed in this paper with $N=46$ nodes. The graph is obtained by superimposing
a RRG sample having $k$ LR links per node (black), onto 
      a one-dimensional ring, where nodes are connected
      by SR links (red). In this example $k=1$.
      Right: graph sample of a generalized RRG  (a portion of) 
     where each node has $k=1$ links (black) of one type,
     playing the role of the LR links of the SW model, plus 2 links of another type (red), playing the role of the SR links
      of the SW-model. In the TDL, where loops become negligible, the two graphs are locally equal.}
    \label{tree}
  \end{center}
\end{figure}

Given a one dimensional ring of $N$ nodes, and a sample of the RRG of degree $k$ superimposed onto the ring, the Hamiltonian of the model (quenched)
with Ising variables $\sigma_i=\pm 1$, reads
\begin{align}
  \label{H}
  H=-J_0\sum_{i=1}^{N}\sigma_i\sigma_{i+1}-J_{}\sum_{(i,j)\in \Gamma}\sigma_i\sigma_j,
\end{align}
where, along the ring, we have assumed periodic boundary conditions, i.e., $\sigma_{N+1}=\sigma_1$,
and $\Gamma$ stands for the set of the $kN/2$ LR links of the RRG realization, with $N$ assumed even.
As customary for mean-field-like systems (possibly anomalous mean-field), including those observed in SW networks~\cite{Soumen}, 
we assume that, in the TDL, self-averaging takes place, i.e., for $N$ sufficiently large,
standard thermal averages become independent from the specific RRG realization.
Moreover, due to the fact that loops become negligible in the TDL, in the same limit the BP approach becomes exact.
More precisely, taking into account that also the ring can be viewed as a RRG of degree $k_0=2$, we see that
our SW model coincides with a generalized RRG of degree $k+2$ having two types of links, SR and LR,
where the SR mimic the one- dimensional underlying structure of the model, as shown for example in Fig.~\ref{tree}
~\cite{Comment}. As we shall see in the following Sections, this mapping between graphs helps to gain a strong insight.

{Below we describe the BP algorithm. For a simple derivation related to the Ising model see e.g. Ref.~\cite{Mezard_Parisi}, while for a more systematic treatment see
  e.g. Refs.~\cite{Yedidia,Mezard_Montanari} where the BP is seen as a distributed transfer--matrix method; see also Ref.~\cite{Review} for the complex network case;
  we also mention that the BP algorithm can be derived in the framework of infinite probabilistic systems by using the
  Kolmogorov consistency theorem, as shown in Ref. \cite{CTBL}. Note that, when applied to the regular Bethe lattice, the BP algorithm is equivalent to the Bethe-Peierls approximation
discovered many years ago by Bethe and Peierls independently.}

For the most general pair-interaction Hamiltonian, i.e., an Hamiltonian of the following form
\begin{align}
  H'=-\sum_{(i,j)}J_{i,j}\sigma_i\sigma_j,
\end{align}
at temperature $\beta^{-1}$ the BP algorithm is defined as
\begin{align}
  \label{BP}
  \beta u^{(n+1)}_{i\to j}=\tanh^{-1}\left[\tanh(\beta J_{i,j})\tanh\left(\sum_{h\in{\partial i\setminus j} }\beta u^{(n)}_{h\to i}\right)\right],
\end{align}
where $\partial i$ stands for the set of first neighbors of node $i$ and $\{u^{(n)}_{i\to j}\}$ represents a set of
$N^2$ {local fields running in correspondence of the $N^2$ directed links ${i\to j}$, i.e., $u^{(n)}_{i\to j}$ is the effective local
magnetic field which, at time $n$, spin $j$ receives from spin $i$}. According to Eqs. (\ref{BP}), the {local fields}
at time $n+1$ are to be evaluated from those at time $n$. Typically, at least in the ferromagnetic case, and given some non pathological initial conditions,
the BP algorithm is known to converge quite fast to a fixed-point, i.e., to a set of stationary {local fields} $\{u_{i\to j}\}$ which can then be used
to evaluate any thermal average. In particular, the mean magnetization of node $i$ is
\begin{align}
  \label{MAG}
  \media{\sigma_i}=\tanh\left(\sum_{j\in{\partial i}} \beta u_{j\to i}\right),
\end{align}
where  
\begin{align}
  \label{BPFP}
  \beta u_{i\to j}=\tanh^{-1}\left[\tanh(\beta J_{i,j})\tanh\left(\sum_{h\in{\partial i\setminus j} }\beta u_{h\to i}\right)\right].
\end{align}

In the following, we apply Eqs. (\ref{BP}) to the Hamiltonian (\ref{H}).

\section{Applying the BP equations}
\begin{figure}
  \begin{center}
    {\includegraphics[height=8cm,clip]
      {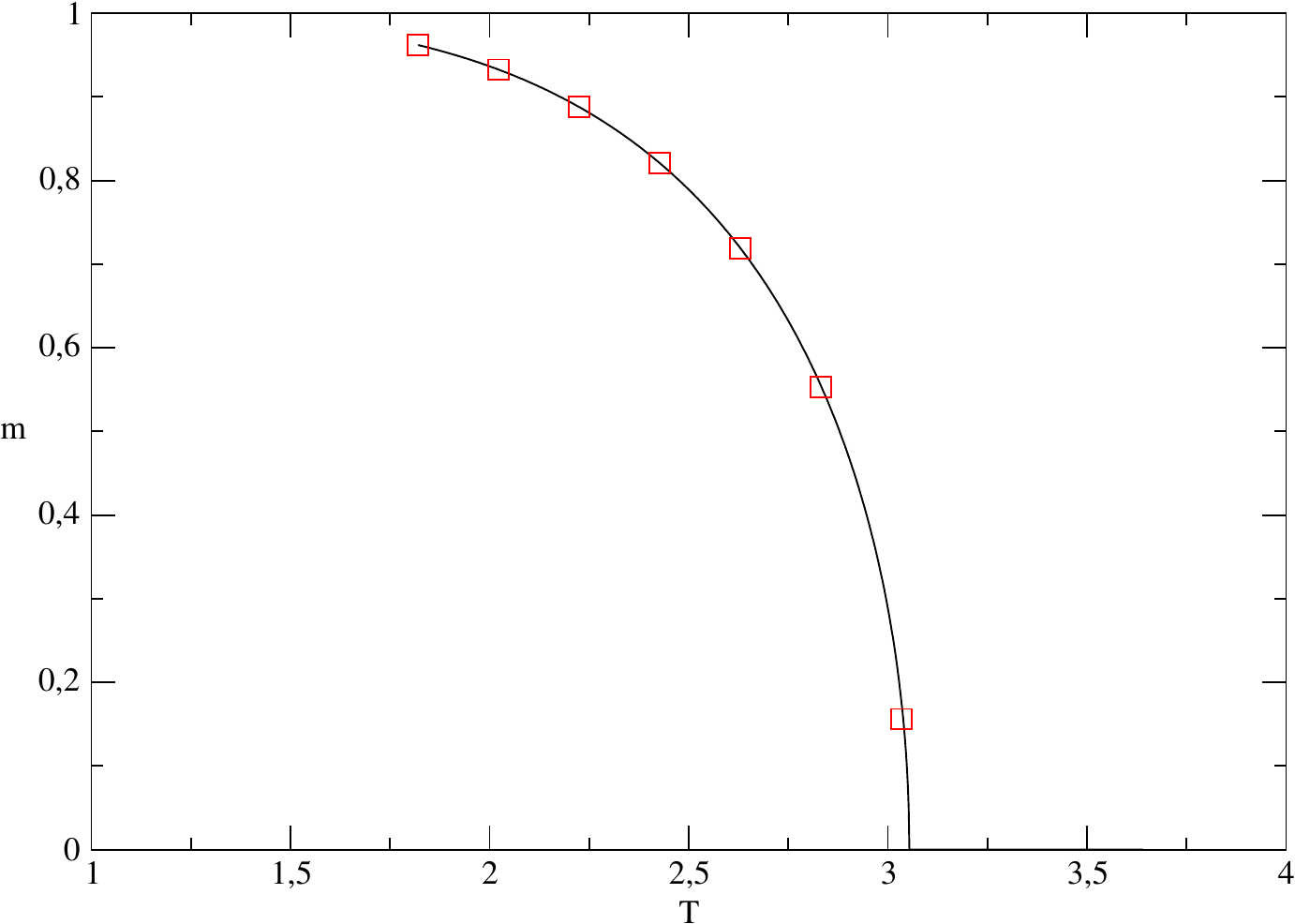}}
    \caption{Plot of the mean magnetization versus temperature obtained as numerical solution of the fixed-point Eqs. (\ref{BPFP1}-\ref{MAG2}) with $k=3$, $J=1$ and $J_0=0.5$. Squares
      are data obtained by Monte Carlo simulations of our SW model having $N=16384$ spins. We have also checked that the iteration of the BP Eqs.~(\ref{BPSW}) with $N=16384$ produces
    a plot indistinguishable from that produced by Eqs. (\ref{BPFP1}-\ref{MAG2}).}
    \label{mag_BP_vs_MC}
  \end{center}
\end{figure}

As we have mentioned in the Introduction, given a $d$-dimensional lattice, there are several ways to transform it into a SW graph via
the addition of LR random links.
In particular, if the desired mean connectivity is $2d+c$, the SW graph can be obtained by connecting $cN/2$ pairs of nodes at random, which results in
a random graph (RG)~\cite{Bollobas,Note2} superimposed onto the $d$-dimensional lattice, or, for $c=k$ integer, we can add the new links at random but in such a way that each node receives
exactly $k$ of these links, which results in a RRG superimposed onto the $d$-dimensional lattice. In both cases, if $d=1$, the resulting SW graph is locally tree-like,
so that the BP approach becomes exact in the TDL.
However, whereas the BP numerical implementation turns out to be quite effective in both cases (at least for ferromagnetic couplings), the RRG SW version
benefits of a fundamental simplification allowing for an analytical approach.
Let us introduce two different symbols for the {local fields} running through a directed LR link (associated to the coupling $J_{}$), denoted as $u_{1;i\to j}$,
and for the {local fields} running through a directed SR link (associated to the coupling $J_0$), denoted as $u_{0;i\to j}$. Now, on observing that
each node has $k$ incoming $u_{1}$-like {local fields}, as well as $2$ incoming $u_0$-like {local fields}, we see that the BP equations (\ref{BP}) can conveniently be rewritten as
\begin{align}
  \label{BPSW}
  \left\{
  \begin{array}{l}
  \beta u^{(n+1)}_{0;i\to j}=\tanh^{-1}\left[t_0\tanh\left(\sum_{h\in \partial_1 i\setminus j}\beta u^{(n)}_{1;h\to i} +\sum_{h\in \partial_0 i\setminus j}\beta u^{(n)}_{0;h\to i} \right)\right],\\
    \beta u^{(n+1)}_{1;i\to j}  =\tanh^{-1}\left[t_{}\tanh\left(\sum_{h\in \partial_1 i\setminus j}\beta u^{(n)}_{1;h\to i} +\sum_{h\in \partial_0 i\setminus j}\beta u^{(n)}_{0;h\to i} \right)\right],
  \end{array}
  \right.
  \end{align}
where $\partial_1 i$ and $\partial_0 i$ stand for the subsets of first neighbors of node $i$ following a LR link and a SR link, respectively
and we have introduced the shorthand notation
\begin{align}
  \label{Notation}
t_0=\tanh(\beta J_0), \quad  t_{}=\tanh(\beta J_{}).
\end{align}
In general, the solution of the fixed-point BP equations (\ref{BPFP}), at least in the ferromagnetic case, tends to follow the underlying graph structure, i.e.,
two {local fields} associated to two links having the same coupling, and topologically equivalent, tend to be equal. However, due to degree fluctuations,
such a topological equivalence is rather absent in a RG or in a SW model based on the RG,
while it is expected in a regular lattice, in the RRG and, partially, in our SW model based on the RRG. More precisely, led by the structure of Eqs. (\ref{BPSW}),
we make the natural ansatz
that the BP equations reach an homogeneous
two-component fixed-point message: one component for the two {local fields} running along the SR links, $u_0$,
and one component for the $k$ {local fields} running along the LR links, $u_{1}$. In other words,
the general fixed-point system (\ref{BPFP}) and mean magnetization simplify as
\begin{align}
  \label{BPFP1}
  & \left\{ 
  \begin{array}{l}
  \beta u_0=\tanh^{-1}\left[t_0\tanh\left(k \beta u_{1} +\beta u_0 \right)\right],\\
   \beta u_{1}  =\tanh^{-1}\left[t_{}\tanh\left((k-1)\beta u_{1} +2\beta u_0 \right)\right],
  \end{array}
  \right. \\
    \label{MAG2}
&  \quad \quad m=\media{\sigma_i}=\tanh\left(k\beta u_{1} +2 \beta u_0\right).
\end{align}
Up  to quadratures, Eqs. (\ref{BPFP1}-\ref{MAG2}) represent the analytic solution of the model (\ref{H}).
In Fig. \ref{mag_BP_vs_MC} we show an example with $k=3$ and compare with Monte Carlo simulations.
Equations (\ref{BPFP1}-\ref{MAG2}) allow to find in a closed form several thermodynamic quantities and, in particular,
the critical temperature, the critical behavior, as well as the correlations near the critical point.
We stress that the above ansatz about the fixed-point holds only for the RRG case while for the RG case, in general, does not hold
due to the fluctuations of the node degrees, preventing therefore a closed analytical approach. More precisely,
in the RG case, as well as in any case where the node degrees are not constant, the fixed-point solutions of the BP equation must be expressed in term of fixed-field-distributions
~\cite{Review} whose solution heavily depends on statistical samplings.
  
\section{Critical point}
\begin{figure}
  \begin{center}
    {\includegraphics[height=8cm,clip]
      {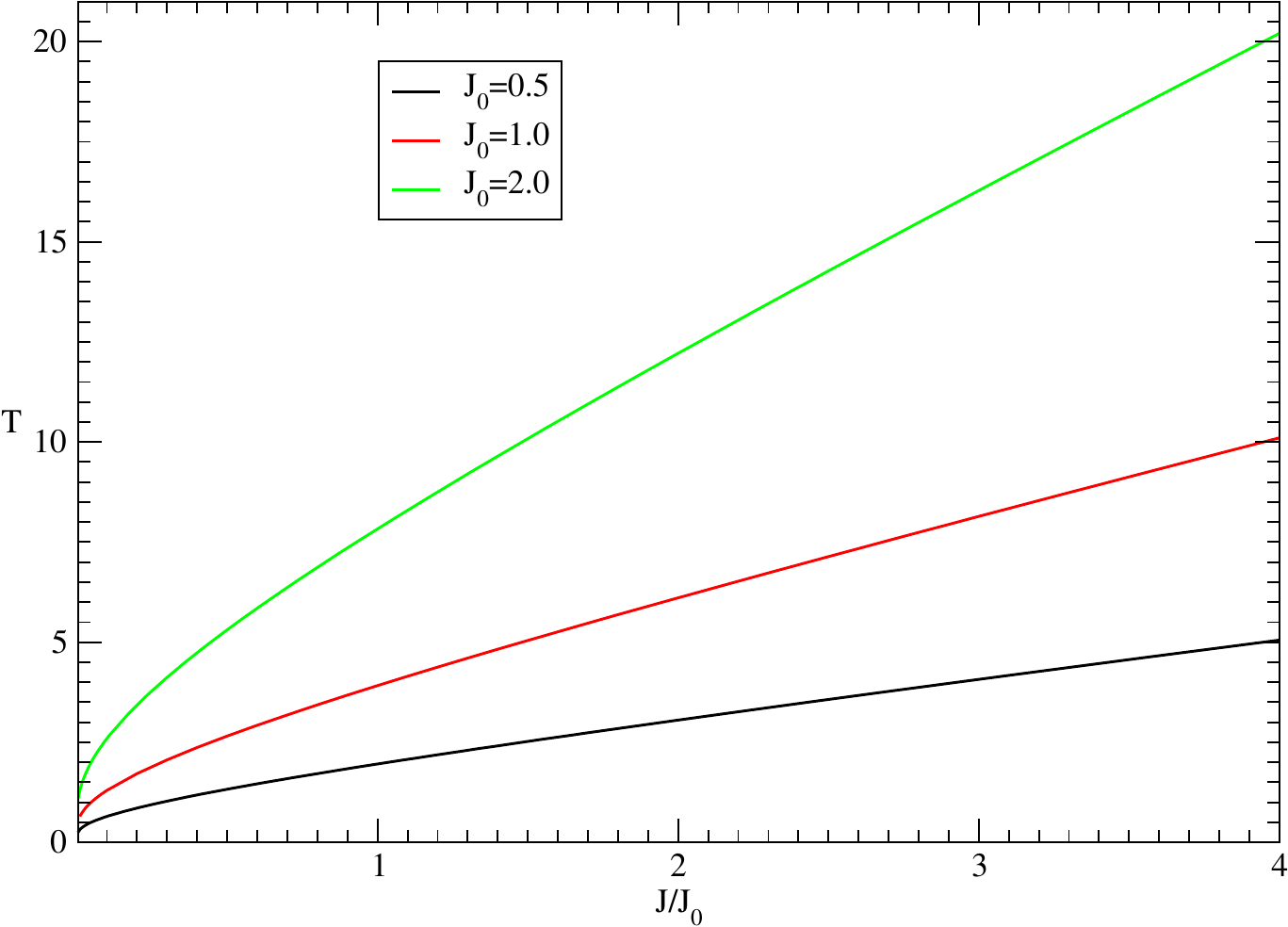}}
    {    \caption{Colors online: Plots of the critical temperature $T_c$ \textit{vs} the ratio $J/J_0$ as numerical solution of Eq. (\ref{Tc}) for three cases with $k=3$.} }
    \label{fig3}
  \end{center}
\end{figure}

On linearizing Eqs. (\ref{BPFP1}), we get the following homogeneous system
\begin{align}
  \label{BPLIN}
  \bm{u}=A\bm{u},
\end{align}
where $\bm{u}$ stands for the two-component vector of the {local fields}
\begin{align}
\label{uvect}
\bm{u}=
\left(
\begin{array}{l}
  u_0\\
  u_{1}
\end{array}
\right),
\end{align}
and $A$ is the following $2\times 2$ matrix
\begin{align}
\label{matrix}
A=
\left(
\begin{array}{l}
  t_0 \quad ~~~ kt_0\\
  2t_{} \quad  (k-1)t_{} \\
\end{array}
\right).
\end{align}
We can now find the critical temperature below which a spontaneous magnetization takes place
by imposing that one of the two eigenvalues of $A$ has modulo equal to 1. Equivalently, and more easily,
if $I$ represents the $2 \times 2$ identity matrix, we can
look for which value of the temperature $\det(A-I)=0$, and look for which value of the temperature (if any) $\det(A+I)=0$, the largest of these two temperature
providing the critical one. For ferromagnetic couplings, it turns out that $\det(A-I)=0$ provides the critical temperature $T_c$ as solution of the following equation
\begin{align}
\label{Tc}
t_{}t_0(k+1)+t_{}(k-1)+t_0=1.
\end{align}
{In Fig. \ref{fig3} we show examples with $k=3$ which makes evident that, for $J\to 0$, $T_c\to 0$ with a vertical slope while, for $J/J_0\to \infty$, the critical temperature
tends to a straight line whose slope increases with $J_0$}.
In the absence of the SR links, i.e. $t_0=0$, Eq. (\ref{Tc}) has a well known structure (see for example~\cite{Review} and references therein)
\begin{align}
\label{TcJ00}
t_{}(k-1)=1, \quad \quad t_0=0,
\end{align}
which gives a finite $T_c$ for any $k\geq 3$.
In the presence of ferromagnetic SR links instead, i.e., $t_0>0$, Eq. (\ref{Tc}) may have a finite $T_c$ even for $k\geq 1$.
Particularly interesting is the symmetric case $t_{}=t_0$, for which Eq. (\ref{Tc}) reduces to
\begin{align}
\label{Tcsymm}
t_{}(k+1)=1, \quad \quad t_{}=t_0.
\end{align}
The resemblance between Eqs. (\ref{TcJ00}) and (\ref{Tcsymm}) is evident and immediately justified: in the symmetric case,
the one-dimensional SW network obtained superimposing a RRG of degree $k$ onto a one-dimensional ring produces a network that, in the TDL, is indistinguishable
from a pure RRG of degree $k+2$, which has a critical temperature whose value satisfies Eq. (\ref{TcJ00}) with $k$ replaced by $k+2$, which
is equivalent to Eq. (\ref{Tcsymm}).

\section{Critical behavior}
In this Section we prove that, as expected, the critical behavior of the system is classical mean-field and show how to determine the related prefactors.
Consider the fixed-point equations.
Above $T_c$ the fixed-point {local fields} are exactly zero hence, from now on, we assume $T<T_c$. Let $\tau$ be the reduced temperature
\begin{align}
\label{Tcsymm2}
\tau=\frac{T_c-T}{T_c}.
\end{align}
Near  the critical point the {local fields} are small so that we can expand in Taylor series the RHS of Eqs. (\ref{BPFP1}). Up to fourth order terms we get
\begin{align}
  \label{CB1}
&  \beta u_0=t_0\left(k \beta u_{1} +\beta u_0 \right)-\frac{t_0(1-t^2_0)}{3}\left(k \beta u_{1} +\beta u_0 \right)^3+\mathop{O}(\| \beta \bm{u}\|^4),\\
  \label{CB2}
&  \beta u_{1}=t^{}\left((k-1) \beta u_{1} +2\beta u_0 \right)-\frac{t(1-t^2)}{3}\left((k-1) \beta u_{1} +2\beta u_0 \right)^3+\mathop{O}(\| \beta \bm{u}\|^4).
\end{align}
Since $\bm{u}\to \bm{0}$ for $\tau \to 0$, we make the following ansatz
\begin{align}
  \label{CBansatz}
  \beta \bm{u}=\bm{a}\tau^{\gamma}, \quad \mathrm{where} \quad \bm{a}=
  \left(
\begin{array}{l}
  a_0\\
  a_1
\end{array}
\right) \quad \mathrm{with} \quad \lim_{T\to T_c}\bm{a}\neq \bm{0},
\end{align}
where the exponent $\gamma>0$, as well as $\bm{a}$, are to be determined self-consistently in the limit $T\to T_c$.
By plugging Eq. (\ref{CBansatz}) into Eqs. (\ref{CB1})-(\ref{CB2}) we get
\begin{align}
  \label{CBa}
(A-I)\bm{a}=\bm{b}\left(\bm{a}\right)\tau^{2\gamma},
\end{align}
where $A$ is given by Eq. (\ref{matrix}), $I$ is the $2\times 2$ identity matrix, and $\bm{b}$ is another vector, function of $\bm{a}$, defined as follows 
\begin{align}
  \label{CBb}
  \bm{b}\left(\bm{a}\right)=
\left(
\begin{array}{l}
  \frac{t_0(1-t^2_0)}{3}\left(k \beta a_{1} +\beta a_0 \right)^3\\
  \frac{t(1-t^2)}{3}\left((k-1) \beta a_{1} +2\beta a_0 \right)^3
\end{array}
\right).
\end{align}
Let us now consider the eigenvalues and eigenvectors of the matrix $A$:
\begin{align}
  \label{CBeigen}
A\bm{v}_i=\lambda_i\bm{v}_i, \quad i=1,2.
\end{align}
Explicitly, for any integer $k$ we have two distinct real eigenvalues 
\begin{align}
  \label{Aeigenv}
\lambda_{1,2}=\frac{kt_0\pm\sqrt{t_0}\sqrt{8kt+t_0(k-2)^2}}{2}.
\end{align}
More precisely, for any integer $k$, $\lambda_1>0$ while $\lambda_2<0$. As a consequence
(and consistently with what we have already seen previously in the section devoted to the critical temperature), we have
$\lim_{T\to T_c}\det (A-I)=0$. In other words, as a function of the temperature $T$, we have $\lim_{T\to T_c}\lambda_{1}(T)-1=0$ while $\lim_{T\to T_c}\lambda_{2}(T)-1\neq 0$.
We shall see in a moment that this fact allows for the determination of both the unknowns in Eq. (\ref{CBansatz}), i.e., the exponent $\gamma$ and the asymptotic value of the vector $\bm{a}$.
Let us consider the Taylor expansion up to a second order term of the parameters $t_0=t_0(T)$ and $t=t(T)$ seen as functions of $T$
\begin{align}
  \label{tt0exp}
  &  t_0(T)=t_{0c}+\beta_c J_0\left(1-t^2_{0c}\right) \tau+\mathop{O}(\tau^2),\\
  &  t(T)=t_{c}+\beta_c J\left(1-t^2_{c}\right)\tau+\mathop{O}(\tau^2),
\end{align}
where $t_{0c}=t_0(T_c)$, $t_{c}=t(T_c)$, and $\beta_c=1/(K_B T_c)$ ($K_B$ being the Boltzmann constant).
By plugging these expressions into $\lambda_1(T)$ in Eq. (\ref{Aeigenv}) we get
\begin{align}
  \label{Aeigenv1}
\lambda_{1}(T)-1=(r_0+r_1)\tau+\mathop{O}(\tau^2), 
\end{align}
where
\begin{align}
  \label{r0}
  &  r_0=\frac{\beta_c J_0}{4}\left(1-t^2_{0c}\right)\left[\frac{2+k t_{0c}}{t_{0c}}+\frac{t_{0c}}{2-k t_{0c}}(k-1^2)\right] \tau+\mathop{O}(\tau^2),\\
    \label{r1}
  &  r_1=\beta_c J\left(1-t^2_{c}\right)\frac{2k t_{0c}}{2-k t_{0c}}\tau+\mathop{O}(\tau^2).
\end{align}
We have now all the ingredients to solve Eq. (\ref{CBansatz}) for the unknown $\bm{a}(T_c)$ and $\gamma$.
In general, as a function of $T$, the vector $\bm{a}(T)$ can be expressed in the eigenbasis of the matrix $A$ as
\begin{align}
  \label{aT}
  \bm{a}(T)=c_1 \bm{v}_1+c_2 \bm{v}_2,
\end{align}
where $c_1$ and $c_2$ are two suitable coefficients and $\|\bm{v}_1\|=\|\bm{v}_2\|=1$.
Note that determining $\bm{a}(T_c)$ amounts to determining $c_1$ and $c_2$ in the limit $T\to T_c$. 
On plugging Eq. (\ref{aT}) into the LHS of Eq. (\ref{CBa}) we obtain
\begin{align}
  \label{CB3}
c_1\left(\lambda_{1}(T)-1\right) \bm{v}_1+c_2\left(\lambda_{2}(T)-1\right) \bm{v}_2=\bm{b}\left(\bm{a}\right)\tau^{2\gamma}.
\end{align}
Now, using the fact that $\lim_{T\to T_c}\lambda_{2}(T_c)-1 \neq 0$, and by using Eqs. (\ref{Aeigenv1})-(\ref{r1}), we see that Eq. (\ref{CB3}) implies the following
\begin{align}
  \label{CB4}
  &  \lim_{T\to T_c}c_2=0,\\
    \label{CB5}
    &  \gamma=\frac{1}{2},\\
      \label{CB6}
  &  \lim_{T\to T_c}c_1\left(r_0+r_1\right)\bm{v}_1=\lim_{T\to T_c}\bm{b}\left(c_1\bm{v}_1\right). 
\end{align}
Up to a quadrature for the third (cubic) equation which allows to determine $c_1$ in the limit $T\to T_c$, Eqs. (\ref{CB4})-(\ref{CB6}) prove that the critical behavior is classical mean-field and yield
the prefactors that, by Taylor expanding Eq. (\ref{MAG2}) up to second order terms, provide the mean magnetization near the critical point as
\begin{align}
    \label{MAG2c}
&  m=\left(ka_{1}(T_c) +2 a_0(T_c)\right)\tau^{1/2}+\mathop{O}(\tau^{3/2}).
\end{align}
In general, the prefactors $a_0(T_c)$ and $a_1(T_c)$ play an irrelevant role. However, they are both proportional to $r_0$ and $r_1$ which,
according to Eqs. (\ref{r0})-(\ref{r1}), have a simultaneous singularity
if the following condition is satisfied  
\begin{align}
    \label{Sing}
& k t_{0c}=2.
\end{align}
On imposing that this condition is satisfied at the critical temperature given by Eq. (\ref{Tc}), we see that such a singularity occurs if and only if
\begin{align}
    \label{Sing2}
& (k^2+k+2)t=k-2,
\end{align}
which has a finite solution for any integer $k\geq 3$. In other words, given an integer $k\geq 3$, if we set $J_0$ such that $k\tanh(J_0/(K_BT_c))=2$ where
$T_c$ is solution of Eq. (\ref{Sing2}), then the prefactors $a_0$ and $a_1$ both diverge in the limit $T\to T_c$. In conclusion, in such a situation, the critical region,
i.e., the temperature interval around $T_c$ where the approximation (\ref{MAG2c}) remains valid, shrinks to zero; a rather interesting feature to be further explored in a future work.

\section{Correlations}
In a locally tree-like graph characterized by a unique coupling $J$,
correlations between two nodes at distance $l$ are known to decay as $t^l$ near the critical point (exactly as $t^l$ above the critical temperature).
However, when there are two different couplings, $J_0$ and $J$, the concept of distance requires a discrimination between the two type of links, LR or SR.
In particular, given two nodes, say $i$ and $j$, they can be first-neighbors to each other in two possible ways. Either they are connected by a LR link associated to the coupling $J$,
or they are connected to each other by a SR link associated to the coupling $J_0$ (there exists also the possibility that the two nodes are connected by the two kinds
of links simultaneously, but this event has probability 0 in the TDL). In order to make this explicit, we shall then indicate the correlations between these two neighbor spins
either as $\media{\sigma_i \sigma_j}_{\mathrm{LR}}$ or as $\media{\sigma_i \sigma_j}_{\mathrm{SR}}$, respectively.
Similarly, given two nodes, $i$ and $j$, they can be second-neighbors to each other in three possible ways: by two LR links, by two SR links, or by one LR followed by one SR link.
It is then clear that such detailed discrimination between nodes at distance $l$ leads to a proliferation of possible distances that grows exponentially with $l$.
For large $l$ would then be more reasonable to consider correlations of specific pairs of nodes regardless of their specific distances.
In this work, however, we will limit ourselves to the analysis of the correlations near the critical point for nodes which are at a specific distance.

Within the BP approach, we can calculate the correlations by using the general method explained for example in \cite{Mezard_Montanari}.
Let $\hat{\nu}_{i\to j}(\sigma)$ be the ``check-to-variable message'' running through the directed pair $i\to j$
(for Hamiltonians involving only pair interactions, the check-to-node link can be shortly replaced with the
corresponding (unique) node-to-node link). 
Note that $\hat{\nu}_{i\to j}(\sigma)$ represents the probability that $\sigma_j=\sigma$ in the modified graph where, expect for the link $i\to j$,
all the other links arriving at $j$ have been removed.
The relation between the check-to-variable messages $\hat{\nu}_{i\to j}(\sigma)$ and the {local fields} previously indicated by $u_{i\to j}$ is simply that the latter correspond to effective
external fields determining the former via the following equality
\begin{align}
    \label{hatnumu}
    & \frac{\hat{\nu}_{i\to j}(-1)}{\hat{\nu}_{i\to j}(+1)}=e^{-2\beta u_{i\to j}}.
\end{align}
At equilibrium, we shall only be interested with the fixed-point messages which, for our one-dimensional SW model, can be of two types:
$\hat{\nu}_{0}(\sigma)$, when the corresponding link is SR, and $\hat{\nu}(\sigma)$, when the corresponding link is LR.

\subsection{Correlation between LR or SR first nearest neighbors}
Let suppose that nodes $i=1$ and $j=2$ are connected by a LR link. Then, the joint probability distribution $\psi_{\mathrm{LR}}(\sigma_1,\sigma_2)$ for spin $\sigma_1$ and $\sigma_2$ reads
\begin{align}
    \label{psiLR}
& \psi_{\mathrm{LR}}(\sigma_1,\sigma_2)\propto e^{\beta J \sigma_1\sigma_2}\hat{\nu}^{2}_{0}(\sigma_1)\hat{\nu}^{k-1}(\sigma_1)\hat{\nu}^{2}_{0}(\sigma_2)\hat{\nu}^{k-1}(\sigma_2),
\end{align}
where the symbol $\propto$ means equality up to a normalization constant. By using Eq. (\ref{hatnumu}) for both $\hat{\nu}_{0}(\sigma)$ and $\hat{\nu}(\sigma)$, we get
the following expression for the correlation function
\begin{align}
    \label{C1LR}
    & \media{\sigma_1 \sigma_2}_{\mathrm{LR}}=\frac
          {(1+t)\left(1+e^{-8\beta u_o-4(k-1)\beta u}\right)-2(1-t)e^{-\beta u_0-2(k-1)\beta u}}
          {(1+t)\left(1+e^{-8\beta u_o-4(k-1)\beta u}\right)+2(1-t)e^{-\beta u_0-2(k-1)\beta u}}.
\end{align}
On Taylor expanding in the {local fields} $u_0$ and $u$ up to second order terms we are left with
\begin{align}
    \label{C1LR2}
    & \media{\sigma_1 \sigma_2}_{\mathrm{LR}}=t+\mathop{O}(\| \beta \bm{u}\|^2).
\end{align}
Let now suppose that nodes $i=1$ and $j=2$ are connected by a SR link. Then, the joint probability distribution $\psi_{\mathrm{SR}}(\sigma_1,\sigma_2)$ for spin $\sigma_1$ and $\sigma_2$ reads
\begin{align}
    \label{psiSR}
& \psi_{\mathrm{SR}}(\sigma_1,\sigma_2)\propto e^{\beta J_0 \sigma_1\sigma_2}\hat{\nu}_{0}(\sigma_1)\hat{\nu}^{k}(\sigma_1)\hat{\nu}_{0}(\sigma_2)\hat{\nu}^{k}(\sigma_2).
\end{align}
By using similar steps as in the previous case we get
\begin{align}
    \label{C1LR2}
    & \media{\sigma_1 \sigma_2}_{\mathrm{SR}}=t_0+\mathop{O}(\| \beta \bm{u}\|^2).
\end{align}

\subsection{Correlation between LR or SR second nearest neighbors}
Evaluating the correlation between second nearest neighbors involves a third spin to be traced out. 
Let suppose that nodes $i=1$, $j=2$ and $k=3$ are such that the pair $(1,2)$ as well as the pair $(2,3)$ are connected by a LR link. Then,
the joint probability distribution $\psi_{\mathrm{LR}}(\sigma_1,\sigma_2,\sigma_3)$ reads
\begin{align}
    \label{psiLR2}
    & \psi_{\mathrm{LR}}(\sigma_1,\sigma_2,\sigma_3)\propto
    e^{\beta J \sigma_1\sigma_2}e^{\beta J \sigma_2\sigma_3}\hat{\nu}^{2}_{0}(\sigma_1)\hat{\nu}^{k-1}(\sigma_1)\hat{\nu}^{2}_{0}(\sigma_2)\hat{\nu}^{k-2}(\sigma_2)\hat{\nu}^{2}_{0}(\sigma_3)\hat{\nu}^{k-1}(\sigma_3),
\end{align}
which leads to
\begin{align}
    \label{C2LR}
    & \media{\sigma_1 \sigma_3}_{\mathrm{LR}}= \frac
    {\sum_{\sigma_1,\sigma_2,\sigma_3}\left(1+t\sigma_1\sigma_2\right)\sigma_1\sigma_3\left(1+t\sigma_2\sigma_3\right)
    \hat{\nu}^{2}_{0}(\sigma_1)\hat{\nu}^{k-1}(\sigma_1)\hat{\nu}^{2}_{0}(\sigma_2)\hat{\nu}^{k-2}(\sigma_2)\hat{\nu}^{2}_{0}(\sigma_3)\hat{\nu}^{k-1}(\sigma_3)}
    {\sum_{\sigma_1,\sigma_2,\sigma_3}\left(1+t\sigma_1\sigma_2\right)\left(1+t\sigma_2\sigma_3\right)
    \hat{\nu}^{2}_{0}(\sigma_1)\hat{\nu}^{k-1}(\sigma_1)\hat{\nu}^{2}_{0}(\sigma_2)\hat{\nu}^{k-2}(\sigma_2)\hat{\nu}^{2}_{0}(\sigma_3)\hat{\nu}^{k-1}(\sigma_3)}.
\end{align}
After some algebra we find
\begin{align}
    \label{C2LR2}
    & \media{\sigma_1 \sigma_3}_{\mathrm{LR}}=t^2+\mathop{O}(\| \beta \bm{u}\|^2).
\end{align}
Similarly we find
\begin{align}
    \label{C2SR2}
    & \media{\sigma_1 \sigma_3}_{\mathrm{SR}}=t_0^2+\mathop{O}(\| \beta \bm{u}\|^2).
\end{align}
and 
\begin{align}
    \label{C1LR1SR}
    & \media{\sigma_1 \sigma_3}_{1,1}=t_0 t+\mathop{O}(\| \beta \bm{u}\|^2),
\end{align}
where $\media{\sigma_1 \sigma_3}_{1,1}$ stands for correlation between two spins separated by two links, one of which is a LR link, while the other is SR link. 

\subsection{Generalization}
Within the BP approach, the above results are exact and have an obvious generalization. If nodes $i$ and $j$ are separated by a chain containing $l=l_{\mathrm{LR}}+l_{\mathrm{SR}}$ links,
$l_{\mathrm{LR}}$ of which are LR links, while the remaining $l_{\mathrm{SR}}$ are SR links,
then their correlation reads
\begin{align}
    \label{CGEN}
    & \media{\sigma_i \sigma_j}_{l_{\mathrm{LR}},l_{\mathrm{SR}}}=t^{l_{\mathrm{SR}}}_0t^{l_{\mathrm{LR}}}+\mathop{O}(\| \beta \bm{u}\|^2)=t^{l_{\mathrm{SR}}}_0t^{l_{\mathrm{LR}}}+\mathop{O}(\tau).
\end{align}
Above the critical temperature, where $\bm{u}=\bm{0}$, Eq. (\ref{CGEN}) reproduces a known result valid for locally tree-like graphs with heterogeneous links.
In particular, the same result applies to a generalized RRG where each node has a specific connectivity for the two types of links.
Note however that here we have proved that, to some extent, such a feature remains robust also below the critical temperature due to the fact that, in all cases, the terms linear in
$\bm{u}$ or, equivalently, the terms linear in the reduced temperature $\tau$, cancel exactly.
In principle, Eq. (\ref{CGEN}) holds for distances at most $\mathop{O}(\log N/\log(k-1))$, beyond which loops must be considered and the BP might lead to wrong results;
moreover, in the presence of loops, more than one path exist between the two nodes so that the concept of specif distance looses meaning.
However, for larger distances correlations
usually becomes negligible and Eq. (\ref{CGEN}) holds effectively.

That the correlations that we have considered in the one-dimensional SW graph look equal to those of a generalized RRG having two types of links have a simple explication:
the two graphs are both locally tree-like and with the same node connectivity containing k LR links plus 2 SR links. See Fig. \ref{tree} for an example with $k=1$. 

\section{Conclusions}
We have analyzed the Ising model built on what could be considered the simplest SW: a RRG superimposed onto a one-dimensional ring.
The resulting SW graph is itself a generalized RRG having two types of links, SR and LR, where the SR ones mimic the underlying one-dimensional ring.
The use of the BP approach in this case is not only exact, but, due to the absence of degree fluctuations, leads also to a closed analytical solution
of the model where we {are} able to derive the equation of state, the critical point, the critical behavior, and the critical correlations. 
We think that the simplicity of this model allows to gain a strong insight about SW models in general. On the other hand, despite this apparent simplicity,
as we have shown in Sec. V, the prefactors characterizing the non universal part of the critical behavior, for certain values of the SR and LR couplings,
own an intriguing singularity, a fact to be further investigated.

\section*{Acknowledgments}
We thank the Dep. of Mathematics of the University of Rome ``La Sapienza'' for hospitality
{(``BANDO PROFESSORI VISITATORI SU PROGETTO DI ECCELLENZA 2023-2027'')}.


%


\section*{References}
\begin{thebibliography}{99}%

\bibitem{Watts} D. J. Watts, S. H. Strogatz, 
\textit{Collective dynamics of 'small-world' networks},
Nature, \textbf{393}, 440 (1998).

\bibitem{Skantzos} N.S. Skantzos and A. C. C. Coolen,
\textit{$(1 + \infty )$-dimensional attractor neural networks},
J. Phys. A: Math. Gen., 5785 \textbf{33} (2000).
  
\bibitem{Barrat} A. Barrat and M. Weigt,
\textit{On the properties of small-world network models},  
Eur. Phys. J B,\textbf{13}, 547 (2000).

\bibitem{Hastings} M. B. Hastings,
\textit{Mean-Field and Anomalous Behavior on a Small-World Network},  
Phys. Rev. Lett. \textbf {91}, 98701 (2003).
  
\bibitem{Viana}  J. Viana Lopes, Yu. G. Pogorelov, J. M. B. Lopes dos Santos, and R. Toral,
\textit{Exact solution of Ising model on a small-world network},  
Phys. Rev. E \textbf{70}, 026112 (2004).

\bibitem{Niko} T. Nikoletopoulos, A. C. C. Coolen, I. Pérez Castillo, N. S. Skantzos, J. P. L. Hatchett and B. Wemmenhove,
\textit{Replicated transfer matrix analysis of Ising spin models on ‘small world’ lattices},  
J. Phys. A: Math. Gen. \textbf{37} 6455 (2004).

\bibitem{Niko2} B. Wemmenhove, T. Nikoletopoulos, and J. P. L. Hatchett,
\textit{Replica symmetry breaking in the ‘small world’ spin glass},  
J. Stat. Mech. P11007 (2005).

\bibitem{Berker} Michael Hinczewski and A. Nihat Berker,
\textit{Inverted Berezinskii-Kosterlitz-Thouless singularity and high-temperature algebraic order
in an Ising model on a scale-free hierarchical-lattice small-world network},  
Phys. Rev. E \textbf{73}, 066126 (2006).  

\bibitem{Hastings2} M. B. Hastings,
\textit{Systematic Series Expansions for Processes on Networks},  
Phys. Rev. Lett. \textbf {96}, 148701 (2006).

\bibitem{SW} 
M. Ostilli and J. F. F. Mendes, 
\textit{Effective field theory for models defined over small-world networks. First- and second-order phase transitions},
Phys. Rev. E \textbf{78}, 031102 (2008).
  
\bibitem{SW2} 
A. L. Ferreira, J. F. F. Mendes, and M. Ostilli, 
\textit{First- and second-order phase transitions in Ising models on small world networks, simulations and comparison with an effective field theory},
Phys. Rev. E \textbf{82}, 011141 (2010).

\bibitem{Bollobas} B. Bollob\'as, \textit{Random Graphs}, Cambridge University Press (2001).

{\bibitem{Note} Sometimes in the literature the RRG is referred to as the ``Bethe Lattice'', although this latter term
should be reserved to indicate a regular tree of fixed degree, which is necessarily an infinite tree, see e.g. Ref.~\cite{CTBL}.}

\bibitem{Mezard_Parisi} M. M\'ezard and G. Parisi,
\textit{The Bethe lattice spin glass revisited},
Eur. Phys. J. B, \textbf{20}, 217 (2001). 

\bibitem{Yedidia} J. S. Yedida, W. T. Freeman, and Y. Weiss,
\textit{Generalized belief propagation},
in Advances in Neural Information Processing Systems, NIPS, pp. 689–695. MIT Press, Cambridge, MA. (2001).

\bibitem{Yedidia2} Yedidia, J. S., Freeman, W. T., and Weiss,
J. S. Yedida, W. T. Freeman, and Y. Weiss,
\textit{Constructing free energy approximations and generalized belief propagation algorithms},
IEEE Trans. Inf. Theory, \textbf{51}, 2282 (2005).

\bibitem{Mezard_Montanari} M. M\'ezard and A. Montanari,
  \textit{Information, Physics, and Computation},
Oxford University Press (2009).  
  
\bibitem{Soumen} R. Soumen and S. M. Bhattacharjee,
\textit{Is small-world network disordered?},
Physics Letters A, \textbf{352}, 13 (2006).

\bibitem{Comment} This kind of argument was also mentioned in a comment in Sec. VI of Ref.~\cite{Viana}.

\bibitem{Note2} Note that, of course, also the RRG is a kind of random graph, however, by the term RG here we mean the classical random graph where only the mean connectivity is fixed, via one 
  two possible RG models, i.e., either the Erd$\ddot{\mathrm{o}}$s-R\'eny random graph (characterized by a total fixed number of links), or the Gilbert model
  (where the total number of links is fixed only on average); see e.g. \cite{Bollobas}.
  
\bibitem{Review} S.N. Dorogovtsev, A.V. Goltsev, J.F.F. Mendes,
\textit{Critical phenomena in complex networks}, 
Rev. Mod. Phys., \textbf{80}, 1275 (2008).


{\bibitem{CTBL} M. Ostilli, \textit{Cayley Trees and Bethe Lattices: A concise analysis for mathematicians and physicists},
Physica A, \textbf{391}, 3417 (2012).}

\end{thebibliography}
\end{document}